\documentclass[%
 reprint,
superscriptaddress,
bibnotes,
 amsmath,amssymb,
 aps,
prb,
]{revtex4-1}

\usepackage{graphicx}% Include figure files
\usepackage{dcolumn}% Align table columns on decimal point
\usepackage{bm}% bold math
\usepackage{color}
\definecolor{OliveGreen}{rgb}{0,0.6,0}
\definecolor{auburn}{rgb}{0.43, 0.21, 0.1}
\definecolor{blue_violet}{rgb}{0.54, 0.17, 0.89}
\definecolor{hokie_maroon}{RGB}{99, 0, 49} % also known as Chicago Maroon
\definecolor{hokie_orange}{RGB}{207, 69, 32} % also known as Burnt Orange

\usepackage{appendix}
\usepackage{chngcntr}
\usepackage{etoolbox}
\usepackage{lipsum}
\usepackage{mathtools}

\AtBeginEnvironment{appendices}{%
\addcontentsline{toc}{section}{Appendices}
\counterwithin{figure}{section}
\counterwithin{equation}{section}
\counterwithin{table}{section}
}

\begin{document}

%\preprint{APS/123-QED}

\title{Application of Cluster Variation and Path Probability Methods to the Tetragonal--Cubic Phase Transition in ZrO$_2$}% Force line breaks with \\

\author{Ryo Yamada}
\email{ryamada@imr.tohoku.ac.jp}
\affiliation{Institute for Materials Research, Tohoku University, Sendai 980-8577, Japan}
\author{Tetsuo Mohri}
\email{tmohri@imr.tohoku.ac.jp}
\affiliation{Institute for Materials Research, Tohoku University, Sendai 980-8577, Japan}

\date{\today}

\begin{abstract}
Cluster variation method (CVM) and path probability method (PPM) have generally been employed to study replacive phase transitions in alloy systems. Recently, displacive phase transitions have been explored within the realm of replacive phase transition in the CVM theoretical framework by viewing displaced atoms as different atomic species, i.e., by converting a freedom of atomic displacement to a configurational freedom. The same methodology is applied to the PPM calculations in this work, and the kinetics of displacive phase transition from tetragonal to cubic phases in ZrO$_2$ are investigated as well as their equilibrium states. 
\end{abstract}

\pacs{Valid PACS appear here}% PACS, the Physics and Astronomy
                             % Classification Scheme.
\maketitle

\section{\label{sec:level1}Introduction}
Phase transition is an important and interesting phenomenon in materials science. There are a variety of phase transitions in metallic alloy systems (e.g., magnetic, configurational, and structural phase transitions), and they have been broadly investigated using various computational approaches, such as molecular dynamics simulation, Monte Carlo method, phase field model, and other statistical mechanics methods, such as the cluster variation method (CVM) \cite{kikuchi1951theory} which is introduced below.

In the statistical mechanics approach, phase stability is analyzed from the free energy, which is generally composed of energy and entropy terms; for example, Helmholtz free energy is defined as $F=E-TS$, where $E$ is the energy, $S$ is the entropy, and $T$ is the temperature. The CVM is one of the most reliable approaches to formulate configurational entropy by considering a broad range of short-range correlations between different atomic species. The CVM has traditionally been combined with an electronic band structure calculation, or density functional theory (DFT), to determine various materials'€™ properties at finite temperatures without using any experimental data (which is the so-called first-principles CVM \cite{mohri2013first}).

Although a rigid lattice is assumed in conventional CVM calculations ignoring local atomic displacements, the continuous-displacement CVM (CDCVM) \cite{kikuchi1998space} enables incorporating local atomic displacements by introducing ``quasi-lattice points''€™ around a Bravais lattice point and viewing the atoms displaced into the quasi-lattice point as different atomic species. The idea of CDCVM (i.e., conversion from a freedom of local atomic displacements to configurational freedom) has been extended to magnetic freedoms \cite{yamada2019conversion} and collective atomic displacements \cite{mohri2013first,kiyokane2018modelling}. In the latter application, for example, the displacive phase transition between tetragonal and cubic phases in ZrO$_2$ is investigated within the scope of replacive transition by regarding upward- and downward-shifted oxygen atoms as different atomic species located on a cubic lattice point.   

This idea of conversion of freedom has also been applied to the path probability method (PPM) \cite{kikuchi1966path}, which is a natural extension of the CVM to the time domain, by the authors of this work \cite{yamada2019atomistic}. Although the main atomic migration process in alloy systems is by a vacancy mechanism, a spin-flipping mechanism (or a direct-exchange mechanism) is generally assumed in the PPM because of the huge computational burden of the vacancy mechanism. By treating various kinetic paths in the PPM as cluster probabilities in the CVM (i.e., by converting a freedom of kinetic paths to a configurational freedom), the computational cost in the PPM calculation is significantly reduced, and an atomic relaxation process in Ni$_3$Al ordered phase was successfully explored by explicitly considering the vacancy mechanism \cite{yamada2019atomistic}. 

In the present study, the kinetics from tetragonal to cubic phase in ZrO$_2$ as well as their equilibrium states are explored within the CVM and PPM frameworks using the aforementioned methodologies; displacive phase transition is studied within the realm of replacive transition, and path variables in the PPM are treated as cluster probabilities in the CVM. This paper is organized as follows. The methodologies of CVM and PPM are, respectively, described in Secs.\;\ref{sec:level2_1} and \ref{sec:level2_2}. The calculated equilibrium states and relaxation paths are shown and discussed in Sec.\;\ref{sec:level3}. Finally, results of tetragonal--cubic phase transitions in ZrO$_2$ using CVM and PPM calculations with CDCVM are summarized in Sec.\;\ref{sec:level4}.

\section{\label{sec:level2}Theory}

\subsection{\label{sec:level2_1}Cluster Variation Method}
The tetragonal--cubic transition in ZrO$_2$ has been investigated using the first-principles CVM by one of the authors in reference \cite{mohri2013first}, where upward- and downward-shifted oxygen atoms along the tetragonal axis are viewed as different atomic species (which are hereafter denoted as A and B atoms, respectively) with fixed zirconium positions. In this methodology, while the tetragonal phase can be described by alternatively displacing oxygen atoms in opposite directions, the cubic phase is realized when the periodicity of the oxygen displacements disappears; thus, this displacive transition can be calculated within the theoretical framework of replacive transitions, more specifically as an order--disorder phase transition \cite{mohri2013first,kiyokane2018modelling}. The same methodology and calculation procedure are followed in the present work, and only the essential points are briefly described below.

\begin{figure}
\begin{center}
\includegraphics[scale=0.48]{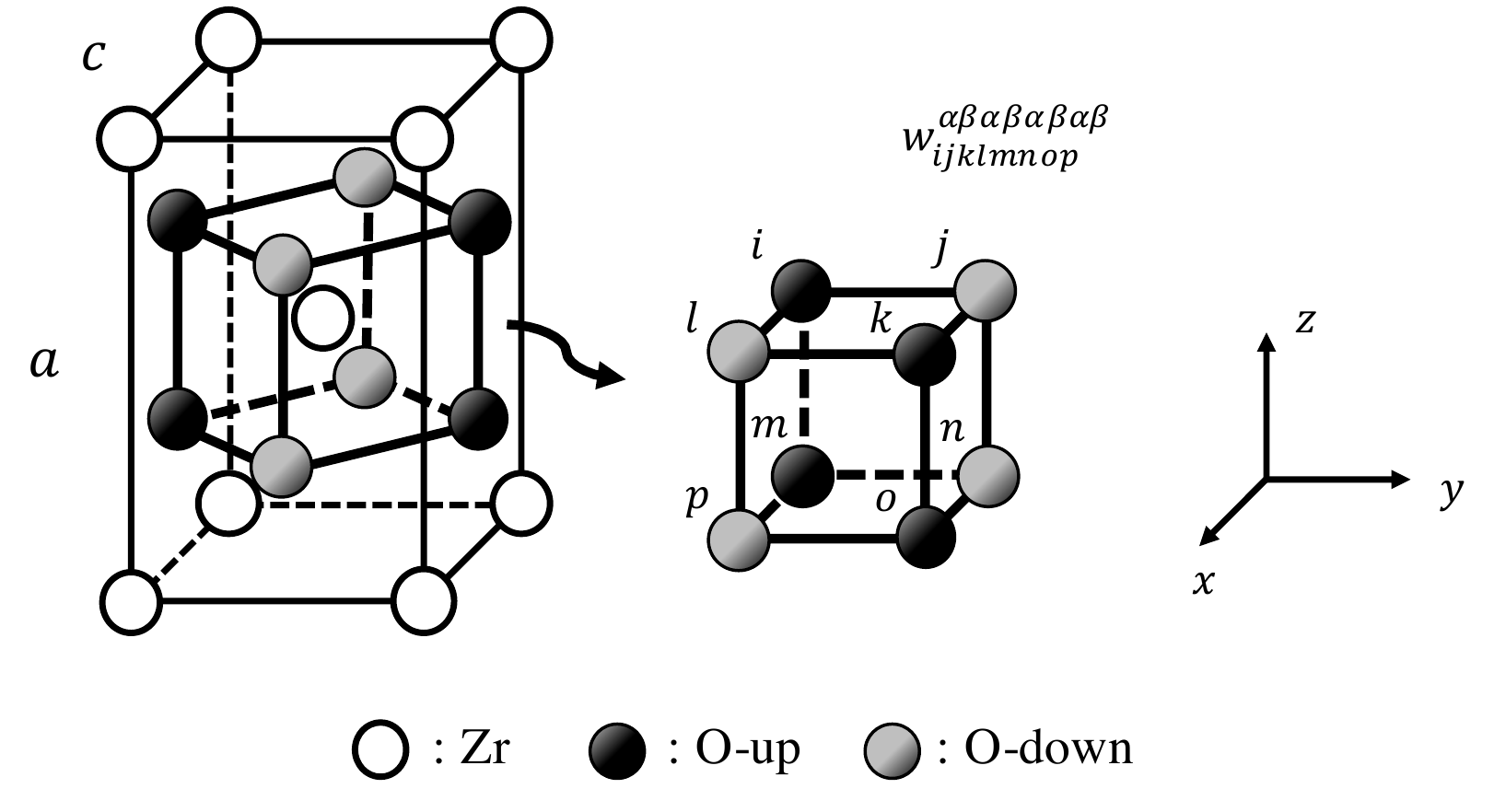}
\caption{\label{fig:sublattice}Cubic structure composed of oxygen atoms in an environment of Zr atoms \cite{mohri2013first}. The tetragonal phase can be described when $\alpha$ and $\beta$ sub-lattices are respectively occupied by up- (A) and down-shifted (B) oxygen atoms. The positions of oxygen atoms in the cubic structure, namely, $i$, $j$, $k$, $l$, $m$, $n$, $o$, and $p$, correspond to the subscripts of the cubic cluster probability $w^{\alpha \beta \alpha \beta \alpha \beta \alpha \beta}_{ijklmnop}$; for example, $w^{\alpha \beta \alpha \beta \alpha \beta \alpha \beta}_{ABABABAB}=1$ for the perfect tetragonal phase. }
\end{center}
\end{figure}

\begin{figure}
\begin{center}
\includegraphics[scale=0.5]{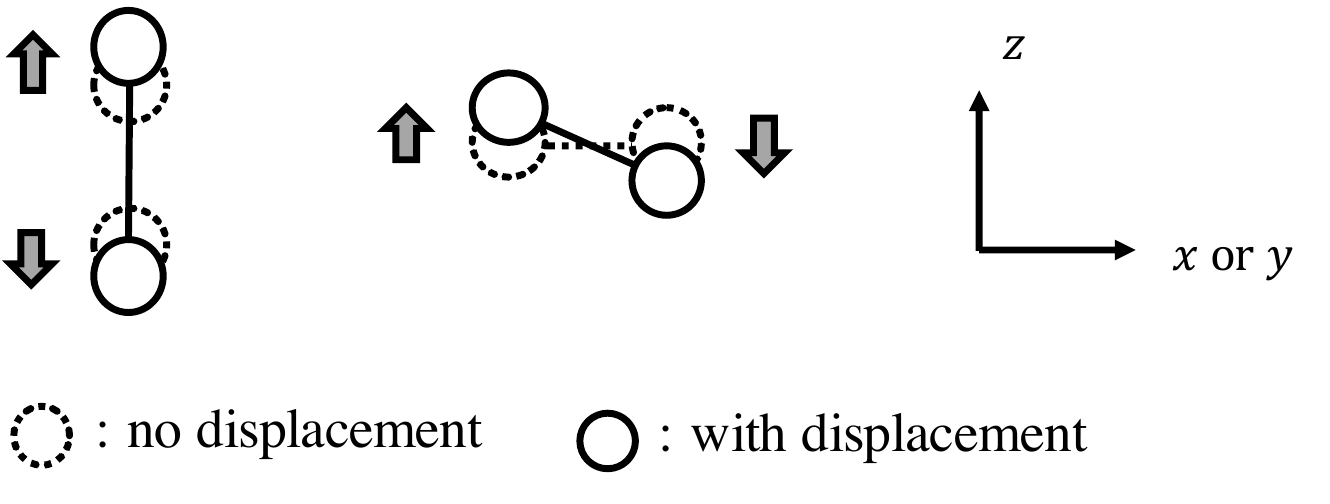}
\caption{\label{fig:atomic_displacement_sublattice}Two types of displacements of up- and down-shifted oxygen atoms depending on the sub-lattices. The left one is on the $\alpha$--$\alpha$ (or $\beta$--$\beta$) sub-lattices and the right one is on the $\alpha$--$\beta$ (or $\beta$--$\alpha$) sub-lattices (see Fig.\;\ref{fig:sublattice}). }
\end{center}
\end{figure}

By considering a cubic structure composed of O atoms in the environment of Zr atoms (see Fig.\;\ref{fig:sublattice}), the total energy, $E$, of the structure is given within the 1st-nearest-neighbor model as
\begin{equation}
\begin{split}
E & =\frac{1}{2}NZ\sum_{i,j}e_{ij} y_{ij} \\
& =\frac{1}{2} N \sum_{i,j} ( e^{\alpha\alpha}_{ij} y^{\alpha\alpha}_{ij} + 4 e^{\alpha\beta}_{ij} y^{\alpha\beta}_{ij} + e^{\beta\beta}_{ij} y^{\beta\beta}_{ij})    \; , \label{eq:total_energy}
\end{split}
\end{equation}
where $N$ is the number of oxygen atoms, $Z$ is the coordination number ($Z=6$ for the cubic structure), $e^{\alpha\alpha/\alpha\beta/\beta\beta}_{ij}$ and $y^{\alpha\alpha/\alpha\beta/\beta\beta}_{ij}$ are, respectively, the 1st-nearest-neighbor pair interaction energies and pair cluster probabilities between $i$ and $j$ atoms in the $\alpha$--$\alpha$/$\alpha$--$\beta$/$\beta$--$\beta$ sub-lattice (see Fig.\;\ref{fig:sublattice}). In the conventional CVM framework, the pair interaction energies, $e^{\alpha\alpha/\alpha\beta/\beta\beta}_{ij}$, are independent of a sub-lattice. In the displacive calculations, however, they may depend on the sub-lattice. In the case considered here, for example, the pair interaction energy between A (up-shifted) and B (down-shifted) atoms is different between $\alpha$--$\alpha$ (or $\beta$--$\beta$) and $\alpha$--$\beta$  (or $\beta$--$\alpha$) sub-lattices, because the atomic displacements lead to different types of displacements (see Fig.\;\ref{fig:atomic_displacement_sublattice}). To distinguish the pair interaction energies, $e_{AB}$ depending on the sub-lattices, they are, respectively, represented by $e^{z}_{AB}$ and $e^{x,y}_{AB}$ (see Table\;\ref{table:pair_interaction_sublattice}). This difference in the pair energy was not taken into account in the preliminary work \cite{mohri2013first}, and the calculated tetragonal--cubic transition temperature of approximately 1500\;K, was significantly underestimated compared with that of the experimental data, which is approximately 2570\;K \cite{aldebert1985structure}. The calculated transition temperature is improved using the pair interaction energies shown in Table\;\ref{table:pair_interaction_sublattice}, as shown in Sec.\;\ref{sec:level3}.

The pair interaction energies, $e_{AA}$, $e^z_{AB}$, and $e^{x,y}_{AB}$, are extracted from the total energies of the three different oxygen arrangements calculated from the full potential linear augmented plane-wave (FLAPW) \cite{jansen1984total}, which are shown in Fig.\;3 in reference \cite{mohri2013first}. The extracted pair interaction energies are fitted to a Lennard--Jones type potential:   
\begin{equation}
e_{ij}(r)= \frac{a_{ij}}{r^7} - \frac{b_{ij}}{r^{3.5}}+c_{ij}    \; ,
\label{eq:L_J_potential}
\end{equation}
where $a_{ij}$, $b_{ij}$, and $c_{ij}$ are the fitting parameters, and $r$ is the interatomic distance. The fitting parameters are shown in Table\;\ref{table:L_J_fitting_parameter}.

\begin{table}
\begin{center}
\caption{\label{table:pair_interaction_sublattice}Pair interaction energies in each sub-lattice $e^{\alpha\alpha/\alpha\beta/\beta\beta}_{ij}$ using $e_{AA}$, $e^{z}_{AB}$, and $e^{x,y}_{AB}$, where $x$, $y$, and $z$ denote the $x$, $y$, and $z$ axes, respectively.  }
\begin{tabular}{ c  c  c  }
$\quad$ & $\quad$ & $\quad$   \\   \hline \hline
  & $\alpha$--$\alpha$ (or $\beta$--$\beta$) \;\; & \;\; $\alpha$--$\beta$  (or $\beta$--$\alpha$)   \\   \hline 
$\quad$ $e_{AA}$ ($=e_{BB}$) $\quad$ & $\quad$ $e_{AA}$ $\quad$ & $\quad$ $e_{AA}$ $\quad$  \\
$e_{AB}$ ($=e_{BA}$)  & $e^{z}_{AB}$ & $e^{x,y}_{AB}$  \\ \hline \hline
\end{tabular}
\end{center}
\end{table}

\begin{table}
\begin{center}
\caption{\label{table:L_J_fitting_parameter}Fitting parameters of Lennard--Jones potentials for each pair. }
\begin{tabular}{ c  c  c  c }
$\quad$ & $\quad$ & $\quad$ & $\quad$ \\   \hline \hline
$\quad\quad$  &  \quad $a_{ij} $ \quad &  \quad $b_{ij} $ \quad  &  \quad $c_{ij} $ \quad   \\ 
$\quad\quad$  &  \; $(\mbox{Ry} \cdot \mbox{au}^7/\mbox{atom}) $ \; & \; $(\mbox{Ry} \cdot \mbox{au}^{3.5}/\mbox{atom}) $ \; & \; $(\mbox{Ry}/\mbox{atom}) $  \\ \hline
\; $e_{AA}$ &  $3.579 \times 10^6$ & $ 2.461  \times 10^3$ & $ 4.226  \times 10^{-1}$ \\
\; $e^z_{AB}$ & $3.599  \times 10^6$ & $2.190  \times 10^3$ & $ 4.277 \times 10^{-1}$  \\
\; $e^{x,y}_{AB}$  &  $3.629  \times 10^6$ & $2.493  \times 10^3$ & $ 4.227 \times 10^{-1}$ \\ \hline \hline
\end{tabular}
\end{center}
\end{table}

The entropy formula for a cubic approximation in the CVM is given by \cite{kiyokane2010order}
\begin{equation}
\begin{split}
S=N k_B \; &\mbox{ln} \Bigg[ \frac{1}{2}\left( \sum_i L(x^{\alpha}_i) + L(x^{\beta}_i) \right) \\
 - \frac{1}{2} & \left( \sum_{i,j} L(y^{\alpha \alpha}_{ij}) + 4  \sum_{i,j} L(y^{\alpha \beta}_{ij}) +  \sum_{i,j} L(y^{\beta \beta}_{ij}) \right)   \\
& + \left( \sum_{i,j,k,l} L(z^{\alpha \beta \alpha \beta}_{ijkl}) + 2 \sum_{i,j,k,l} L(z^{\alpha \alpha \beta \beta}_{ijkl})  \right)  \\
& \;\;\;\;\;\;\;\;\;\;\; - \sum_{i,j,k,l,m,n,o,p} L(w^{\alpha \beta \alpha \beta \alpha \beta \alpha \beta}_{ijklmnop})  \Bigg] 
\; , \label{eq:entropy_cubic}
\end{split}
\end{equation}
where $k_B$ is the Boltzmann constant, $x^{\alpha/\beta}_i$, $y^{\alpha\alpha/\alpha\beta/\beta\beta}_{ij}$, $z^{\alpha \beta \alpha \beta/\alpha \alpha \beta \beta}_{ijkl}$, and $w^{\alpha \beta \alpha \beta \alpha \beta \alpha \beta}_{ijklmnop}$ are the respective cluster probabilities of the point, pair, square, and cubic clusters at the sub-lattices, and $L(x) \equiv x\;\mbox{ln}x -x$.

The free energy of a system, $F=E-TS$, is formulated using Eqs.\;\eqref{eq:total_energy} and \eqref{eq:entropy_cubic}. An equilibrium state of a system is determined by minimizing the free energy with respect to the cluster probabilities and interatomic distances:
\begin{equation}
\left( \frac{\partial F}{\partial w^{\alpha \beta \alpha \beta \alpha \beta \alpha \beta}_{ijklmnop}} \right) = 0 \;\;\;\; \mbox{and} \;\;\;  \left( \frac{\partial F}{\partial r} \right) = 0
  \; . \label{eq:free_energy_minimization}
\end{equation}
The minimization is achieved by the natural iteration method (NIM) \cite{kikuchi1974superposition}.

\subsection{\label{sec:level2_2}Path Probability Method}
The huge computational burden of the PPM can be significantly reduced by treating the path variables as cluster probabilities in the CVM. Whereas the vacancy mechanism is required for an atomic diffusion process, as in reference \cite{yamada2019atomistic}, an atomic displacement can be modeled within the mold of the spin-flipping mechanism. 

\begin{table}
\begin{center}
\caption{\label{table:path_variables}Path variables defined in this work: (a) point, (b) pair, (c) square, and (d) cubic path variables. Some representative path variables are shown for the pair, square, and cubic cases, and the superscripts of the sub-lattices are omitted. Note that the point path variables are numbered as 1, 2, 3, and 4. Subsequently, the path variables of pair, square, and cubic cases are expressed by these numbers. }
\includegraphics[scale=0.43]{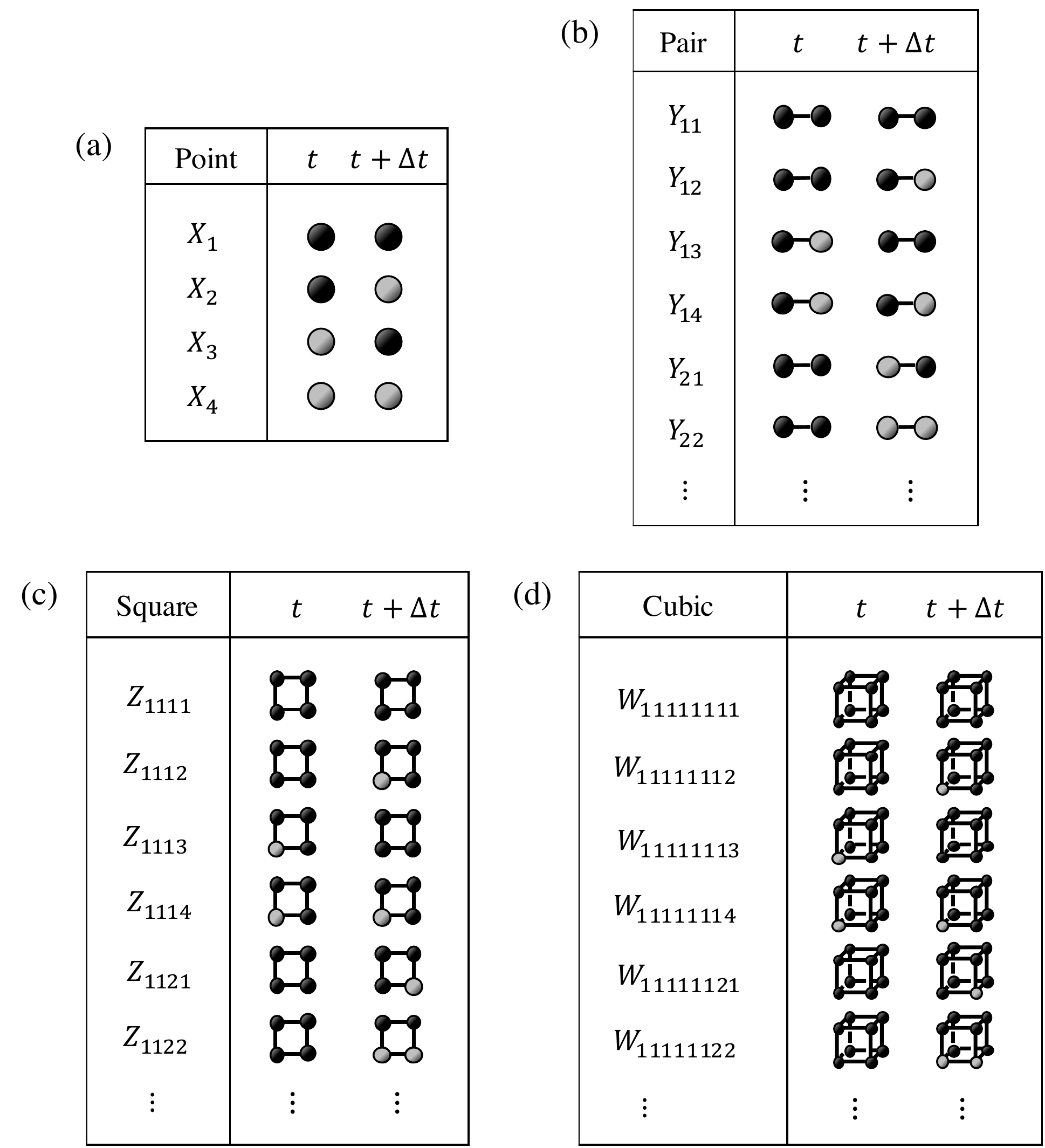}
\end{center}
\end{table}

The PPM framework is based on the assumption that a system takes the most probable kinetic path during a relaxation process. The most probable path is determined by maximizing a path probability function (PPF). The PPF, $P$, is a product of three terms; $P=P_1 P_2 P_3$, where $P_1$ is the probability of spin-flipping, $P_2$ is the probability of thermal activation for a state change, and $P_3$ is the number of equivalent paths. The logarithmic expressions of these three terms are written as 
\begin{equation}
\begin{split}
\mbox{ln} P_1& = \frac{1}{2} N  \Big[  \mbox{ln} \left( \theta \cdot \Delta t \right) \left( \left( X^\alpha_2 + X^\alpha_3 \right) + \left( X^\beta_2 + X^\beta_3 \right) \right) \\
& +  \mbox{ln} \left( 1- \theta \cdot \Delta t \right) \left( \left( X^\alpha_1 + X^\alpha_4 \right) + \left( X^\beta_1 + X^\beta_4 \right) \right)  \Big]
  \; ,  \label{eq:P1_term}
\end{split}
\end{equation}
\begin{equation}
\mbox{ln} P_2=  -\frac{\Delta E_{\mbox{\scriptsize{system}}}}{2k_B T}  \; ,  \label{eq:P2_term}
\end{equation}
and
\begin{equation}
\begin{split}
\mbox{ln} P_3 & = N \Bigg[ \frac{1}{2}\left( \sum_i L(X^{\alpha}_i) + L(X^{\beta}_i) \right) \\
 & - \frac{1}{2}  \left( \sum_{i,j} L(Y^{\alpha \alpha}_{ij}) + 4  \sum_{i,j} L(Y^{\alpha \beta}_{ij}) +  \sum_{i,j} L(Y^{\beta \beta}_{ij}) \right)   \\
& \;\;\;\; + \left( \sum_{i,j,k,l} L(Z^{\alpha \beta \alpha \beta}_{ijkl}) + 2 \sum_{i,j,k,l} L(Z^{\alpha \alpha \beta \beta}_{ijkl})  \right)  \\
& \;\;\;\;\;\;\;\;\;\;\;\;\;\;\; - \sum_{i,j,k,l,m,n,o,p} L(W^{\alpha \beta \alpha \beta \alpha \beta \alpha \beta}_{ijklmnop})  \Bigg] 
\; , \label{eq:P3_term}
\end{split}
\end{equation}
where $\theta$ is the spin-flipping probability, $\Delta t$ is the time step, $X^{\alpha/\beta}_i$, $Y^{\alpha\alpha/\alpha\beta/\beta\beta}_{ij}$, $Z^{\alpha \beta \alpha \beta/\alpha \alpha \beta \beta}_{ijkl}$, and $W^{\alpha \beta \alpha \beta \alpha \beta \alpha \beta}_{ijklmnop}$ are the respective path variables of the point, pair, square, and cubic clusters in the sub-lattices, which are defined in Table\;\ref{table:path_variables}, and $\Delta E_{\mbox{\scriptsize{system}}}$ is a change in internal energy before and after spin-flippings in $\Delta t$. The spin-flipping frequency, $\theta$, and internal energy change, $\Delta E_{\mbox{\scriptsize{system}}}$, are written as follows:
\begin{equation}
\theta = \theta_0 \cdot \mbox{exp}\left(  -\frac{Q}{k_B T} \right)  \; \label{eq:theta_detail}
\end{equation}
and
\begin{equation}
\begin{split}
& \Delta E_{\mbox{\scriptsize{system}}} = \frac{1}{2} N \Big[ \sum_{i,j} e^{\alpha\alpha}_{ij} \left( y^{\alpha\alpha}_{ij} (t+\Delta t) - y^{\alpha\alpha}_{ij} (t)  \right) \\
& \;\;\;\;\;\;\;\;\;\;\;\;\;\;\;\;\;\;\; + 4 \sum_{i,j} e^{\alpha\beta}_{ij} \left( y^{\alpha\beta}_{ij} (t+\Delta t) - y^{\alpha\beta}_{ij} (t)  \right) \\
& \;\;\;\;\;\;\;\;\;\;\;\;\;\;\;\;\;\;\;\;\;\;\;\;\;\;\; + \sum_{i,j} e^{\beta\beta}_{ij} \left( y^{\beta\beta}_{ij} (t+\Delta t) - y^{\beta\beta}_{ij} (t)  \right)  \Big] \\
& =  \frac{1}{2} N \left[ \sum_{i,j} \Delta e^{\alpha\alpha}_{ij} Y^{\alpha\alpha}_{ij} + 4 \sum_{i,j} \Delta e^{\alpha\beta}_{ij} Y^{\alpha\beta}_{ij} + \sum_{i,j} \Delta e^{\beta\beta}_{ij} Y^{\beta\beta}_{ij} \right]
  , \label{eq:deltaE_system_detail}
\end{split}
\end{equation}
where $\theta_0$ is the attempt frequency, $Q$ is the activation energy, and $\Delta e^{\alpha\alpha/\alpha\beta/\beta\beta}_{ij}$ are the energy changes of pair interaction energies in $\Delta t$. The maximization of the PPF is performed by the NIM \cite{ohno2005iteration} as well as minimization of the free energy in the CVM in Sec.\;\ref{sec:level2_1}. 

Note that while the subscripts of cluster probabilities and pair interaction energies indicate atomic species (e.g., $x_A$ and $e_{AA}$), those of path variables and the change of pair interaction energies represent a type of kinetic path (e.g., $X_1$ and $\Delta e_{11}$). This indicates that the path variables are treated in the same way as the cluster variables in CVM. \cite{yamada2019atomistic}

\section{\label{sec:level3}Results and Discussion}
\begin{figure}
\begin{center}
\includegraphics[scale=0.52]{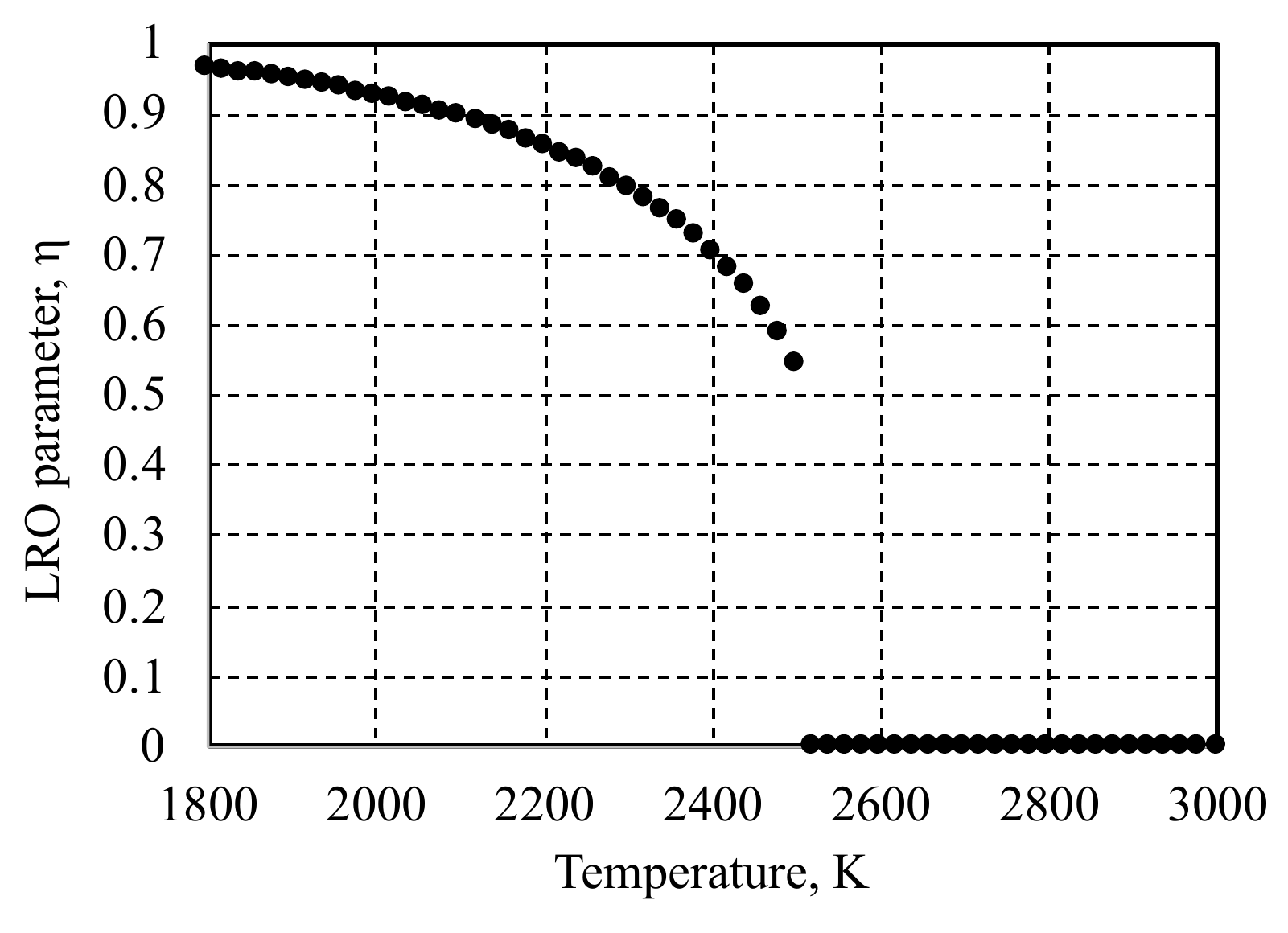}
\caption{\label{fig:temperature_dependence_LRO}Temperature dependence of the LRO parameter calculated by CVM. }
\end{center}
\end{figure}

The degree of tetragonality compared to a cubic phase is evaluated using the long-range order (LRO) parameter, $\eta$, which is defined as \cite{mohri2013first}
\begin{equation}
\eta = \frac{x^\alpha_A - x^\alpha_B}{2}  \; . \label{eq:long_range_order_definition}
\end{equation}
The LRO parameters at the equilibrium states are shown in Fig.\;\ref{fig:temperature_dependence_LRO}. As can be seen from the results, the transition temperature is estimated to be around 2500\;K. This is quite close to the experimental value of 2570\;K, unlike the transition temperature reported in reference \cite{mohri2013first}, where appropriate pair interaction energies are not used for the energy term (see Sec.\;\ref{sec:level2_1}). 

It is controversial whether the tetragonal--cubic phase transition in ZrO$_2$ is first-order or second-order \cite{fabris2001free}. Although the transition is considered to be weakly second-order in the preliminary study \cite{mohri2013first}, the calculated temperature dependence of the LRO parameter shown in Fig.\;\ref{fig:temperature_dependence_LRO} represents a first-order characteristic, i.e., it discontinuously becomes zero. In the preliminary and present calculations, however, only the displacements of oxygen atoms are considered with the fixed tetragonality, $c/a$, of the lattice composed of zirconium atoms (see Fig.\;\ref{fig:sublattice}). It has been noted that the degree of tetragonality also depends on the temperature and plays a role in the mechanism of the transition \cite{fabris2001free}. Therefore, a more detailed analysis is required for a determination of the order of the tetragonal--cubic phase transition in ZrO$_2$ within the CVM framework.

\begin{figure}
\begin{center}
\includegraphics[scale=0.64]{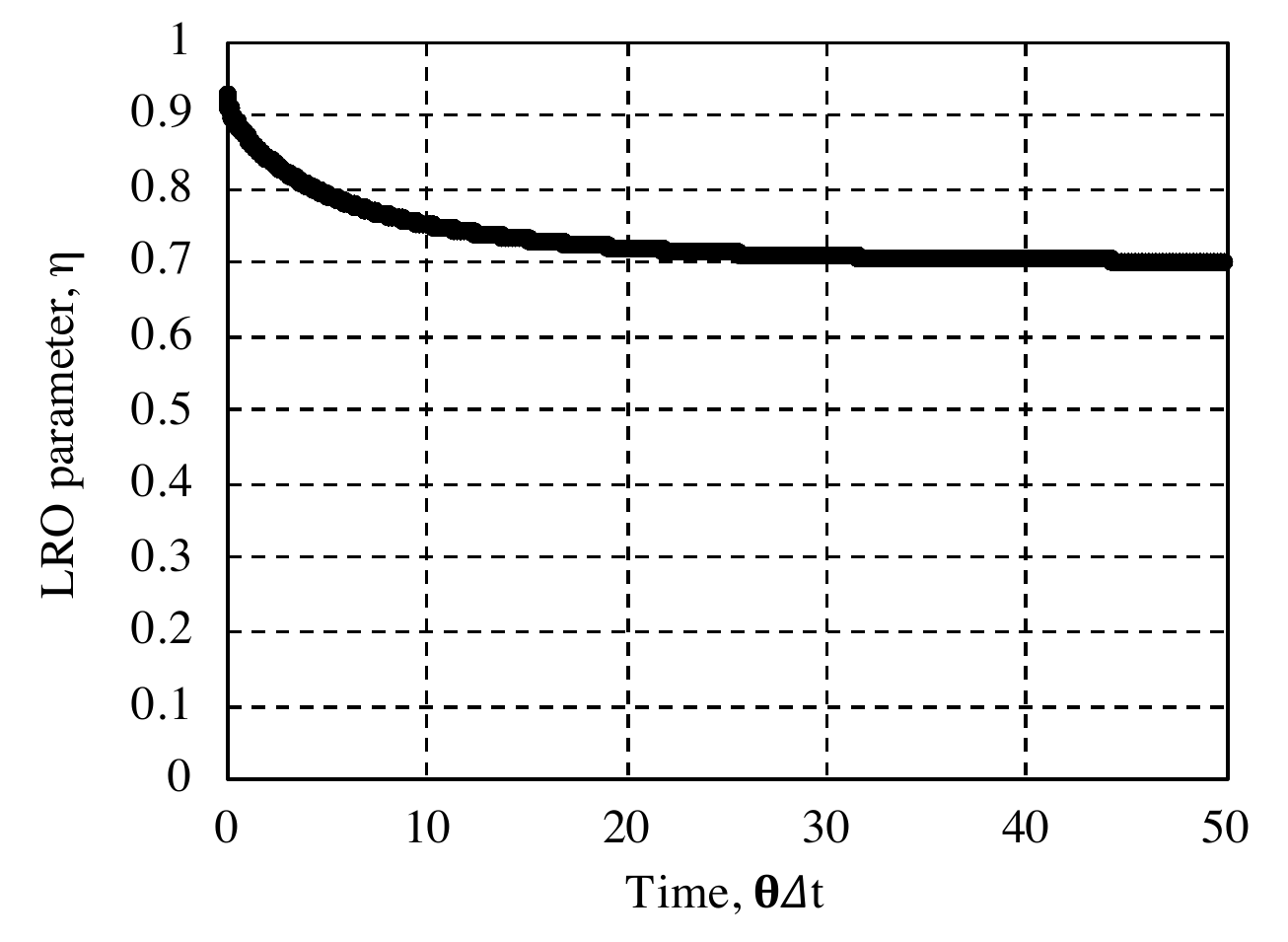}
\caption{\label{fig:relaxation_2400}Time dependence of the LRO parameter calculated by PPM when a sample is quenched from $T_0=2000$\;K to $T_R=2400$\;K. }
\end{center}
\end{figure}

The relaxation process for a sample quenched from $T_0=2000$\;K to $T_R=2400$\;K is shown in Fig.\;\ref{fig:relaxation_2400}. It can be seen that the change of the LRO parameter takes place at the initial stage of the relaxation process and the final equilibrium state corresponds to the one independently calculated from the CVM, $\eta \approx 0.7$ (see Fig.\;\ref{fig:temperature_dependence_LRO}). 

Note that the time scales here are normalized by the spin-flipping frequency, $\theta$. Their real time dependence can be determined once the attempt frequency, $\theta_0$, and the activation energy, $Q$, are calculated from the electronic band structure calculation, but this is beyond the scope of the present study and will be considered in the future work.

\section{\label{sec:level4}Conclusions}
The displacive transitions from tetragonal to cubic phases in ZrO$_2$ are investigated using the CVM and PPM. In both methodologies, the idea of CDCVM is employed; i.e., displacive and kinetic freedoms are converted into configurational freedoms. The transition temperature from tetragonal to cubic phase is estimated to be around 2500\;K, which is quite close to the experimental data, 2570\;K, and the relaxation process is successfully calculated using the PPM.   
 
 One of the biggest advantages of the PPM over other kinetic models (e.g., kinetic Monte Carlo and phase field model) is that the calculated relaxation process can be easily scaled using the data determined from an electronic band structure calculation. It has been shown that there is a certain relation between relaxation processes calculated from PPM and phase field model \cite{ohno2006critical}. Thus, it would be possible to incorporate a real time dependence into the phase field simulations by effectively combining with the PPM calculation.

\section*{ACKNOWLEDGEMENT}
This study is partly supported by a project (P16010) commissioned by the New Energy and Industrial Technology Development Organization (NEDO), and by the Structural Materials for Innovation of the Cross ministerial Strategic Innovation Promotion Program (SIP) of Japan Science and Technology (JST).

\bibliographystyle{ieeetr}
\bibliography{ref}

\end{document}